# Low-loss Lithium Niobate on Insulator (LNOI) Waveguides of a 10 cm-length and a Sub-nanometer Surface Roughness

Rongbo Wu [1,2,†], Min Wang [3,4,†], Jian Xu [3,4], Jia Qi [1,2], Wei Chu [1,4], Zhiwei Fang [3,4], Jianhao Zhang [1,2], Junxia Zhou [3,4], Lingling Qiao [1], Zhifang Chai [3,4], Jintian Lin [1], and Ya Cheng [1,3,4,5,*]

1. State Key Laboratory of High Field Laser Physics, Shanghai Institute of Optics and Fine Mechanics, Chinese Academy of Sciences, Shanghai 201800, China; rbwu@siom.ac.cn (R.W.); qijia@siom.ac.cn (J.Q.); chuwei0818@qq.com (W.C.); jhzhang@siom.ac.cn (J.Z.); qiaolingling@siom.ac.cn (L.Q.); jintianlin@siom.ac.cn (J.L.)
2. University of Chinese Academy of Sciences, Beijing 100049, China;
3. State Key Laboratory of Precision Spectroscopy, East China Normal University, Shanghai 200062, China; mwang@phy.ecnu.edu.cn (M.W.); jxu@phy.ecnu.edu.cn (J.X.); zwfang@phy.ecnu.edu.cn (Z.F.); 1780497299@qq.com (J.Z.); zfchai@phy.ecnu.edu.cn (Z.C.)
4. XXL—The Extreme Optoelectromechanics Laboratory, School of Physics and Materials Science, East China Normal University, Shanghai 200241, China
5. Collaborative Innovation Center of Extreme Optics, Shanxi University, Taiyuan, Shanxi 030006, China

* Correspondence: ya.cheng@siom.ac.cn (Y.C.)
† These authors contributed equally to this paper.

**Abstract:** We develop a technique for realizing lithium niobate on insulator (LNOI) waveguides of a multi-centimeter-length with a propagation loss as low as 0.027 dB/cm. Our technique relies on patterning a chromium (Cr) thin film coated on the top surface of LNOI into a hard mask with a femtosecond laser followed by the chemo-mechanical polishing for structuring the LNOI into the waveguides. The surface roughness on the waveguides is determined to be 0.452 nm with an atomic force microscope (AFM). The approach is compatible with other surface patterning technologies such as optical and electron beam lithographies or laser direct writing, enabling high-throughput manufacturing of large-scale LNOI-based photonic integrated circuits.

## 1. Introduction

Photonic integrated circuits (PICs) have shown promising potential for realizing complex information processing systems employing both quantum and classical light sources [1,2]. To increase the computational efficiency and reconfigurability, the crucial requirements for the PIC-based optical computers/calculators are low propagation loss, fast tunability, and efficient optical interfacing. Currently, several materials have been utilized to construct large-scale PICs, including silicon and some semiconductor materials [3-6], fused silica [7,8] and bulk lithium niobate (LN) [9,10]. The advantage of silicon-based PICs is the high refractive index of silicon which can allow for fabricating compact light circuits with strong confinement and tight bends. In addition, the lithographic technology for high precision patterning of silicon and semiconductors is mature. However, silicon-based PICs intrinsically suffer from a relatively high propagation loss and a transmission window prohibitive for visible and shorter wavelengths. PICs can be built on fused silica and bulk LN crystals by local modification of the refractive index with either illumination of light or ion doping. Unfortunately, the refractive index increases achieved using these approaches are usually on the orders of $10^{-3}$ to $10^{-2}$, resulting in large footprints of the PICs as required for minimizing the bending loss. Most importantly, the typical propagation losses of waveguides in the state-of-the-art PICs are typically on the order of $10^{-1}$ dB/cm or higher, which sets an ultimate limitation on the performance of the PIC-based optical computers.

Recently, a revolutionary approach for building high performance PICs has been emerging enabled by the successful demonstration of high quality lithium niobate on insulator (LNOI) nanophotonic structures. The first experimental proof of this approach is provided by first patterning the LNOI into the designated geometries using a femtosecond laser. The draft structures obtained after the femtosecond laser patterning, which has a relatively high sidewall roughness on the order of tens of nanometers, are then polished with a focused ion beam (FIB) milling to smoothen the sidewall [11]. This concept was soon extended to incorporate with other lithographic technologies such as optical lithography and electron beam writing (EBW) for defining the planar patterns on LNOI substrates followed by reactive ion etching to complete the nanostructuring of the LNOI [12,13]. The initial focus was mainly placed on optical microresonators [11-21], and other devices such as waveguides and photonic crystals [22-28] appeared shortly, taking the advantage of high surface smoothness of the sidewalls as a result of the ion dry etching. So far, the propagation loss in the LNOI waveguides has reached 0.04 dB/cm which opens the avenue toward large-scale PIC applications [29].

It is noteworthy that the ion etching step which is necessary for achieving the high quality sidewalls on the LNOI nanophotonic structures inherently leaves a low but non-negligible surface roughness which is difficult to be completely removed [29]. Moreover, the use of FIB or EBW in the patterning of LNOI makes the approach impractical for fabricating large-scale PICs owing to their low throughputs and limited ranges of motion. Recently, we have developed a technique for fabricating high quality optical microresonators on LNOI with a quality factor above $10^7$ [30]. Since this technique does not involve any ion beam etching processes, surface smoothness beyond that allowed by ion beam etching can be readily achieved, and the footprint sizes of PICs can be greatly increased by patterning the LNOI photonic structures with either laser direct writing or optical lithography. Here, we experimentally show that we are able to realize 10 cm-long LNOI waveguides of a propagation loss of 0.027 dB/cm which is benefited from the low surface roughness of 0.452 nm measured with an atomic force microscope (AFM). The low loss waveguides can be essential building blocks for light modulation, beam delivery and manipulation, nonlinear optics, and optical signal processing.

**2. Materials and Methods**

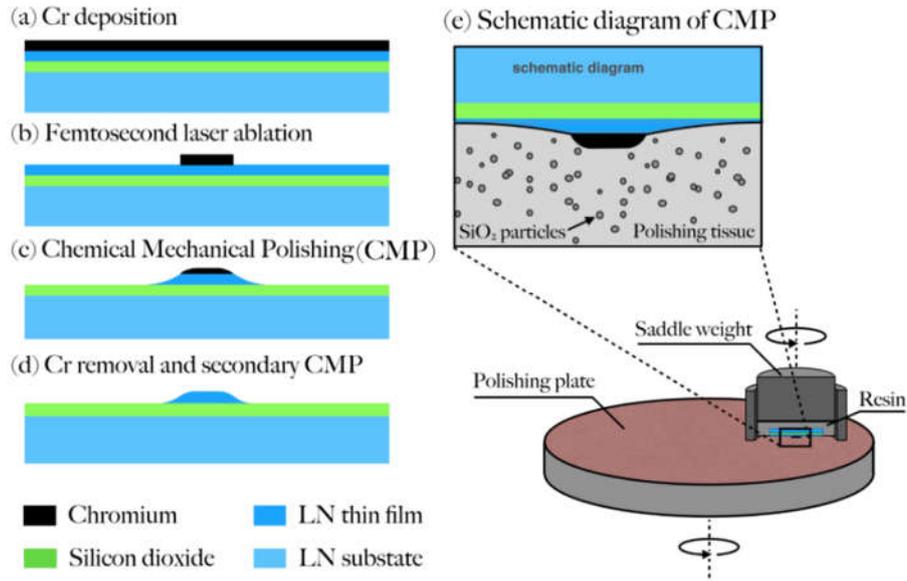

**Figure 1.** (**a**)-(**d**) Flow-chart of fabrication of lithium niobate on insulator (LNOI) waveguide and (**e**) schematic diagram of chemo-mechanical polishing (CMP).

In our experiment, the LNOI waveguides were produced on a commercially available X-cut LNOI wafer fabricated by ion slicing (NANOLN, Jinan Jingzheng Electronics Co., Ltd.) [31]. The LN thin film with a thickness of 400 nm is bonded to a 2 μm thick $SiO_2$ layer grown on a LN substrate. The fabrication process includes four steps, as schematically illustrated in Figure 1. First, a thin layer of chromium (Cr) with a thickness of 600 nm was deposited on the surface of the LNOI by magnetron sputtering. Subsequently, the patterning of Cr film on the LNOI sample was carried out using femtosecond laser ablation. Specifically, to minimize the heat effect as well as the redeposition of the ablation debris on the surface, the femtosecond laser ablation was conducted by focusing the femtosecond laser beam onto the top surface of LNOI sample which was immersed in water. The key in this step is to carefully choose a pulse energy of the femtosecond laser so as to enable a complete removal of the Cr film without damaging the underneath LNOI thanks to the high precision and low heat generation in the interaction of femtosecond laser pulses with various types of materials [32]. The detailed parameters of femtosecond laser ablation are provided as follows. We used a femtosecond laser with a center wavelength of 1030 nm, a pulse width of 170 fs, and a tunable repetition rate ranging from 2.5 kHz to 83.3 kHz (PHAROS, LIGHT CONVERSION) for the Cr film patterning. A variable neutral density filter was used for adjusting the average power. A 100× objective lens (M Plan Apo NIR, Mitutoyo Corporation) with a numerical aperture (NA) of 0.7 was used to produce a tightly focused spot of ~ 1 μm diameter. The LNOI sample was mounted on a computer-controlled XY motion stage (ABL15020WB and ABL15020, Aerotech) with a translation resolution of 100 nm, whereas positioning of the focus of laser beam in the Z direction was achieved with another one-dimensional motion stage of a translation resolution of 100 nm (ANT130-110-L-ZS, Aerotech) onto which the objective lens was mounted. A charged coupled device (CCD) connecting with the computer was installed above the objective lens to monitor the fabrication process in real time. The laser power was chosen to be 0.2 mW at which a complete removal of the Cr thin film can be realized, while the LNOI underneath the Cr film remains intact for its relatively higher damage threshold than Cr. The remaining Cr disk formed on the LNOI serves as a hard mask for the subsequent CM polishing.

The CM polishing process was conducted using a wafer polishing machine (NUIPOL802, Kejing, Inc.). In the CM polishing process, we used a piece of velvet polishing cloth, and the polishing slurry (MasterMet, 60 nm amorphous colloidal silica suspension) was provided by

Buehler, Ltd. The soft velvet cloth allows not only the Cr film but also the exposed LNOI to be accessed by the polishing slurry. Since the Cr film is of a higher hardness than that of the LNOI, the exposed LNOI could be preferentially removed in the CM polishing process before the sacrifice of the Cr mask. Moreover, by carefully controlling the duration of the CM polishing process, the side wall angle with respect to the vertical direction of the fabricated LNOI nanostructures can be controlled from a few degrees for a long polishing duration to nearly 80° for a short duration of the polishing process. Finally, the fabricated structure was first immersed in a Cr etching solution (Chromium etchant, Alfa Aesar GmbH) for 10 min to remove the Cr mask, and then polished again at a relatively low polishing pressure with a shorter polishing duration to improve the smoothness of the upper surface of LN waveguide.

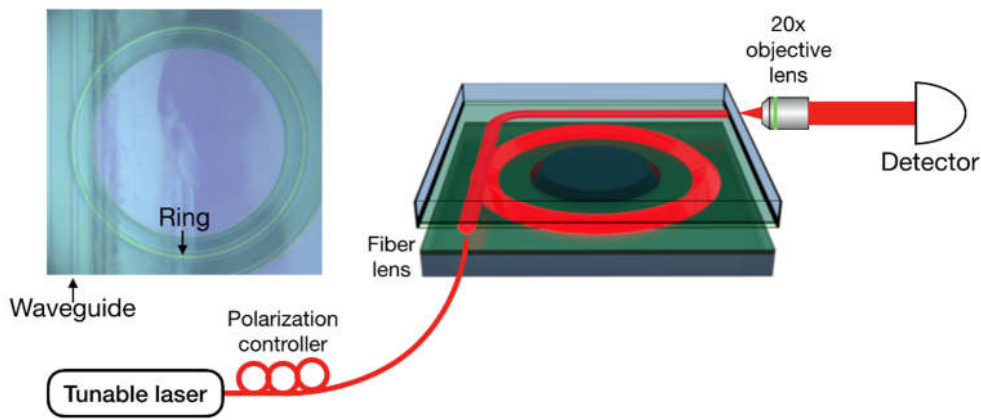

**Figure 2.** Schematic of the experimental setup for measuring the Q factor of the microring resonator. Left inset: Optical micrograph image of the microring resonator coupling with the waveguide, as indicated by the black arrows.

To characterize the propagation loss in the LNOI waveguide, we constructed a whispering gallery ring resonator by which the propagation loss of the LNOI waveguide can be determined using $\alpha = 2\pi n_{eff} / (Q\lambda)$, where $\alpha$ is the attenuation coefficient, $n_{eff}$ the effective refractive index, Q the quality factor of the ring resonator, and $\lambda$ the wavelength of the light beam. Both the $n_{eff}$ and the Q-factor can be determined from the transmission spectrum of microring resonator. The experimental setup for measuring the Q factor of the ring resonator is schematically shown in Figure 2. The light produced by a tunable laser (LTB-6728, New Focus) was coupled into a curved LNOI waveguide whose geometric parameters including the thickness and width are the same as that of the LNOI ring by use of a fiber lens. The exiting light was collected by a 20× objective lens (MPlanFL N, OLYMPUS) into a detector (Model 1811, New Focus). The bend of the upper waveguide was intentionally introduced for preventing the stray light from the fiber laser from entering the objective lens located in front of the detector. The polarization direction of the light was adjusted with a fiber polarization controller. The curved waveguide was coupled to the LNOI ring resonator by evanescent coupling. Specifically, by carefully adjusting the distance between the coupling waveguide and the microring resonator, we are able to achieve the critical coupling condition which is crucial for obtaining an accurate intrinsic Q factor of the ring resonator.

**3. Results**

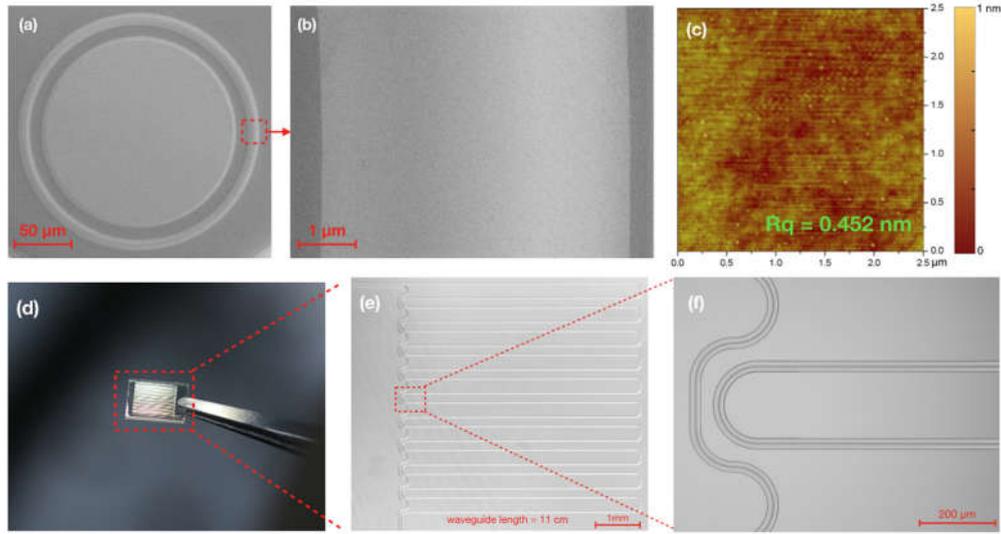

**Figure 3.** (**a**) Top-view scanning electron microscope (SEM) image of a LN microring resonator. (**b**) Zoomed view of the ridge of the microring resonator in (a). (**c**) Atomic force microscope (AFM) image of the ridge. (**d**) Picture of a chip consisting of an 11 cm-long waveguide captured by digital camera. (**e**)-(**f**) Zoomed images of the waveguides on the chip captured by an optical microscope.

Figure 3a shows the top-view scanning electron microscope (SEM) images of a LN micro ring resonator with a diameter of 160 μm and the width of the waveguide is ~3 μm. The close-up-view SEM of an arc of the ring highlights the high surface smoothness on the CM polished sample. A further atomic force microscope (AFM) inspection as illustrated in Figure 3c confirms that a surface roughness as low as $R_q$= 0.452 nm has been achieved. The same fabrication technique was also used to fabricate a continuous 11 cm-long optical waveguide as shown by the digital-camera-captured picture in Figure 3d, with the details being given by the zoomed optical micrographs in Figure 3e - f. The total time of femtosecond laser ablation for fabricating the 11 cm-long waveguide is 90 min. At this moment, the footprint size of the PICs is only limited by the LNOI wafer size. Technically speaking, the LNOI wafer can be made larger without much difficulty [33].

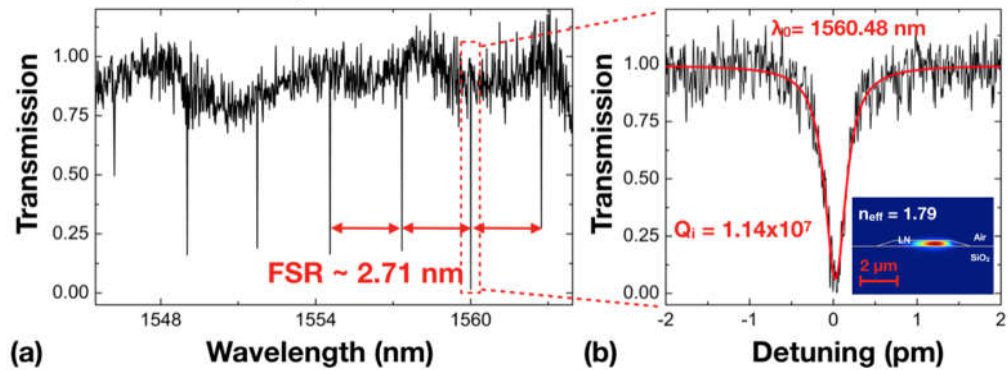

**Figure 4.** (**a**) Transmission spectrum of the LN microring resonator. (**b**) The Lorentz fitting (red curve) reveals a loaded Q factor of $5.70 \times 10^6$, corresponding to an intrinsic Q factor of $1.14 \times 10^7$. Inset: The optical mode distribution and $n_{eff}$ in the ring waveguide calculated using FDTD simulation.

The optical loss characterization was performed using whispering-gallery-resonator-loss measurements. The propagation loss $\alpha$ is related to Q-factor of the ring resonator as mentioned before. Figure 4a shows the measured transmission spectrum for the wavelength range between 1546 nm and 1564 nm. The free spectral range (FSR) of the microring resonator is determined to be 2.71 nm, which is consistent with the diameter of our ring resonator of 160 μm. The resonant lines appear regularly spaced, indicating that mostly only the low-order modes exist in the ring resonator. One of the whispering-gallery modes at the resonant wavelength of 1560.48 nm is chosen for the measurement of the loaded Q-factor by fitting with a Lorentz function, which reached $5.07 \times 10^6$, corresponding to an intrinsic Q-factor of $1.14 \times 10^7$ in the critical coupling regime as evidenced by Figure 4b. The effective refractive index $n_{eff} = \lambda^2 / (2\pi R \cdot FSR)$, with the ring radius of R = 80 μm and wavelength of 1560.48 nm, is calculated to be 1.79, which is in good agreement with our FDTD simulation result given in the inset of Figure 4b. Combining the effective refractive index and the Q-factor obtained from the transmission spectrum, the propagation loss in the microring resonator is calculated to be 0.027 dB/cm using the aforementioned expression $\alpha = 2\pi n_{eff} / (Q\lambda)$. This result represents the upper limit of the propagation loss in the LNOI waveguide fabricated using our method.

## 4. Discussion

The low propagation loss of 0.027 dB/cm is a result of the low surface roughness of $R_q$=0.452 nm on the fabricated LNOI waveguides. This is benefitted from our technique in which the LNOI is purely patterned by the chemo-mechanical polishing without any use of the ion beam etching [30]. The ion beam etching inherently leaves a small amount of surface roughness on the nearly vertical sidewalls which is difficult to be completely removed by top surface polishing [34,35]. This is the major reason that we are able to obtain a propagation loss lower than the waveguides fabricated by FIB or reactive ion etching. In the current experiment, we used a femtosecond laser to pattern the Cr hard mask. Generally speaking, the sidewall roughness on the Cr mask patterned by the femtosecond laser ablation should be higher than the surface roughness on the LNOI photonic structures produced with the ion beam etching. However, the sidewall roughness on the Cr mask will only be transferred to the underneath LNOI near the top surface, thus it can be completely suppressed with an additional polishing process for thinning the LNOI substrate after the removal of the Cr mask (see, Figure 1d).

It should be mentioned that the propagation loss obtained by measuring the Q factor of the ring resonator may have been underestimated for the straight segments in the LNOI waveguides as presented in Figure 3d - f due to a higher radiative loss in the ring resonator. Ultimately, the propagation loss in the LNOI waveguides is limited by the absorption in crystalline LN which is well known to be on the order of $\sim 10^{-3}$ dB/cm. Our measured loss is still one order of magnitude away from the theoretical limit, which could be attribute to several likely factors including the radiative loss in the ring resonator, some unknown contamination on the surface of LNOI waveguide as the measurements are all carried out in a low-class clean room, existence of absorptive defects in the LNOI substrate owing to the imperfect crystal growth, and the remaining surface roughness left by the chemo-mechanical polishing process. Thus, to realize LNOI waveguides with propagation losses on the order of $10^{-3}$ dB/cm, a lot of refinements should be systematically investigated in the future.

The fabrication resolution of femtosecond laser direct writing is typically on the order of 1 μm for inorganic materials such as glass, semiconductor and metal. However, today's optical lithography can easily achieve sub-micron or even 100 nm-level patterning resolutions. This should be sufficient for fabricating single mode LNOI waveguides of narrower widths by which PICs such as Mach-Zender interferometers and polarization convertors can be built. Moreover, the mode field of the LNOI waveguides can be tuned by coating them with fused silica for suppressing higher order modes as well as scattering loss thanks to the reduced contrast of the refractive index between the LNOI waveguide and the cladding environment [36]. With all of these improvements, the LNOI waveguides will become a major building block for PIC applications.

## 5. Conclusions

To conclude, we have achieved a propagation loss of 0.027 dB/cm in LNOI waveguides fabricated by a combination of femtosecond laser micromachining for patterning the Cr mask and chemo-mechanical polishing for transferring the laser-written patterns to the LNOI beneath the Cr mask. By eliminating the low-throughput FIB or EBW process, our technique can enable rapid fabrication of low-loss optical waveguides of great lengths which are only limited by the motion range of the air bearing stage and the size of LNOI wafers. Thus, our approach greatly promotes the fabrication efficiency and reduces the cost in the manufacturing of LNOI PICs. The waveguides as well as ring resonators can be adopted for constructing complex PICs, opening the avenue for mass production of LNOI PICs for optical communication and computation applications.

**Conflicts of Interest:** The authors declare no conflict of interest.

## References

1. Ladd, T.D.; Jelezko, F.; Laflamme, R.; Nakamura, Y.; Monroe, C.; O'Brien, J.L. Quantum computers. *Nature* **2010**, *464*, 45, doi:10.1038/nature08812.
2. Shen, Y.; Harris, N.C.; Skirlo, S.; Prabhu, M.; Baehr-Jones, T.; Hochberg, M.; Sun, X.; Zhao, S.; Larochelle, H.; Englund, D., et al. Deep learning with coherent nanophotonic circuits. *Nature Photon.* **2017**, *11*, 441, doi:10.1038/nphoton.2017.93.
3. Harris Nicholas, C.; Bunandar, D.; Pant, M.; Steinbrecher Greg, R.; Mower, J.; Prabhu, M.; Baehr-Jones, T.; Hochberg, M.; Englund, D. Large-scale quantum photonic circuits in silicon. In *Nanophotonics*, 2016; Vol. 5, p 456, doi:10.1515/nanoph-2015-0146.
4. Dietrich, C.P.; Fiore, A.; Thompson, M.G.; Kamp, M.; Höfling, S. GaAs integrated quantum photonics: Towards compact and multi-functional quantum photonic integrated circuits. *Laser Photon. Rev.* **2016**, *10*, 870-894, doi:10.1002/lpor.201500321.
5. Najafi, F.; Mower, J.; Harris, N.C.; Bellei, F.; Dane, A.; Lee, C.; Hu, X.; Kharel, P.; Marsili, F.; Assefa, S., et al. On-chip detection of non-classical light by scalable integration of single-photon detectors. *Nature Comm.* **2015**, *6*, 5873, doi:10.1038/ncomms6873
6. Yang, K.Y.; Oh, D.Y.; Lee, S.H.; Yang, Q.-F.; Yi, X.; Shen, B.; Wang, H.; Vahala, K. Bridging ultrahigh-Q devices and photonic circuits. *Nature Photon.* **2018**, *12*, 297-302, doi:10.1038/s41566-018-0132-5.
7. Qiang, X.; Zhou, X.; Wang, J.; Wilkes, C.M.; Loke, T.; O'Gara, S.; Kling, L.; Marshall, G.D.; Santagati, R.; Ralph, T.C., et al. Large-scale silicon quantum photonics implementing arbitrary two-qubit processing. *Nature Photon.* **2018**, *12*, 534-539, doi:10.1038/s41566-018-0236-y.
8. Marshall, G.D.; Politi, A.; Matthews, J.C.F.; Dekker, P.; Ams, M.; Withford, M.J.; O'Brien, J.L. Laser written waveguide photonic quantum circuits. *Opt. Express* **2009**, *17*, 12546-12554, doi: 10.1364/OE.17.012546.
9. Sohler, W.; Hu, H.; Ricken, R.; Quiring, V.; Vannahme, C.; Herrmann, H.; Büchter, D.; Reza, S.; Grundkötter, W.; Orlov, S., et al. Integrated optical devices in lithium niobate. *Opt. Photon. News* **2008**, *19*, 24-31, doi:10.1364/OPN.19.1.000024.
10. Jin, H.; Liu, F.M.; Xu, P.; Xia, J.L.; Zhong, M.L.; Yuan, Y.; Zhou, J.W.; Gong, Y.X.; Wang, W.; Zhu, S.N. On-chip generation and manipulation of entangled photons based on reconfigurable lithium-niobate waveguide circuits. *Phys. Rev. Lett.* **2014**, *113*, 103601, doi:10.1103/PhysRevLett.113.103601.
11. Lin, J.; Xu, Y.; Fang, Z.; Song, J.; Wang, N.; Qiao, L.; Fang, W.; Cheng, Y. Second harmonic generation in a high-Q lithium niobate microresonator fabricated by femtosecond laser micromachining. *arXiv preprint arXiv:1405.6473* **2014**.
12. Wang, J.; Bo, F.; Wan, S.; Li, W.; Gao, F.; Li, J.; Zhang, G.; Xu, J. High-Q lithium niobate microdisk resonators on a chip for efficient electro-optic modulation. *Opt. Express* **2015**, *23*, 23072-23078, doi: 10.1364/OE.23.023072.

13. Wang, C.; Burek, M.J.; Lin, Z.; Atikian, H.A.; Venkataraman, V.; Huang, I.-C.; Stark, P.; Lončar, M. Integrated high quality factor lithium niobate microdisk resonators. *Opt. Express* **2014**, *22*, 30924-30933, doi: 10.1364/OE.22.030924.
14. Lin, J.; Xu, Y.; Fang, Z.; Wang, M.; Song, J.; Wang, N.; Qiao, L.; Fang, W.; Cheng, Y. Fabrication of high-Q lithium niobate microresonators using femtosecond laser micromachining. *Sci. Rep.* **2015**, *5*, 8072, doi: 10.1038/srep08072.
15. Lin, J.; Xu, Y.; Ni, J.; Wang, M.; Fang, Z.; Qiao, L.; Fang, W.; Cheng, Y. Phase-matched second-harmonic generation in an on-chip $LiNbO_3$ microresonator. *Phys. Rev. Appl.* **2016**, *6*, 014002, doi: 10.1103/PhysRevApplied.6.014002.
16. Lin, J.; Xu, Y.; Fang, Z.; Wang, M.; Wang, N.; Qiao, L.; Fang, W.; Cheng, Y. Second harmonic generation in a high-Q lithium niobate microresonator fabricated by femtosecond laser micromachining. *Sci. China Phys. Mech. Astron.* **2015**, *58*, 114209, doi: 10.1007/s11433-015-5728-x.
17. Hao, Z.; Wang, J.; Ma, S.; Mao, W.; Bo, F.; Gao, F.; Zhang, G.; Xu, J. Sum-frequency generation in on-chip lithium niobate microdisk resonators. *Photon. Res.* **2017**, *5*, 623-628, doi: 10.1364/PRJ.5.000623.
18. Wang, L.; Wang, C.; Wang, J.; Bo, F.; Zhang, M.; Gong, Q.; Lončar, M.; Xiao, Y.-F. High-Q chaotic lithium niobate microdisk cavity. *Optics Lett* **2018**, *43*, 2917-2920, doi:10.1364/OL.43.002917.
19. Luo, R.; Jiang, H.; Rogers, S.; Liang, H.; He, Y.; Lin, Q. On-chip second-harmonic generation and broadband parametric down-conversion in a lithium niobate microresonator. *Opt. Express* **2017**, *25*, 24531-24539, doi: 10.1364/OE.25.024531.
20. Jiang, W.C.; Lin, Q. Chip-scale cavity optomechanics in lithium niobate. *Sci. Rep.* **2016**, *6*, 36920, doi: 10.1038/srep36920.
21. Fang, Z.; Xu, Y.; Wang, M.; Qiao, L.; Lin, J.; Fang, W.; Cheng, Y. Monolithic integration of a lithium niobate microresonator with a free-standing waveguide using femtosecond laser assisted ion beam writing. *Sci. Rep.* **2017**, *7*, 45610, doi: 10.1038/srep45610.
22. Liang, H.; Luo, R.; He, Y.; Jiang, H.; Lin, Q. High-quality lithium niobate photonic crystal nanocavities. *Optica* **2017**, *4*, 1251-1258, doi:10.1364/OPTICA.4.001251.
23. Wang, C.; Zhang, M.; Stern, B.; Lipson, M.; Lončar, M. Nanophotonic lithium niobate electro-optic modulators. *Opt. Express* **2018**, *26*, 1547-1555, doi: 10.1364/OE.26.001547.
24. Wang, C.; Xiong, X.; Andrade, N.; Venkataraman, V.; Ren, X.-F.; Guo, G.-C.; Lončar, M. Second harmonic generation in nano-structured thin-film lithium niobate waveguides. *Opt. Express* **2017**, *25*, 6963-6973, doi:10.1364/OE.25.006963.
25. Luo, R.; He, Y.; Liang, H.; Li, M.; Lin, Q. Highly tunable efficient second-harmonic generation in a lithium niobate nanophotonic waveguide. *Optica* **2018**, *5*, 1006-1011, doi:10.1364/OPTICA.5.001006.
26. Cai, L.; Han, H.; Zhang, S.; Hu, H.; Wang, K. Photonic crystal slab fabricated on the platform of lithium niobate-on-insulator. *Optics Lett* **2014**, *39*, 2094-2096, doi:10.1364/OL.39.002094.
27. Diziain, S.; Geiss, R.; Steinert, M.; Schmidt, C.; Chang, W.-K.; Fasold, S.; Füßel, D.; Chen, Y.-H.; Pertsch, T. Self-suspended micro-resonators patterned in Z-cut lithium niobate membranes. *Opt. Mater. Express* **2015**, *5*, 2081-2089, doi:10.1364/OME.5.002081.
28. Krasnokutska, I.; Tambasco, J.-L.J.; Li, X.; Peruzzo, A. Ultra-low loss photonic circuits in lithium niobate on insulator. *Opt. Express* **2018**, *26*, 897-904, doi:10.1364/OE.26.000897.
29. Wolf, R.; Breunig, I.; Zappe, H.; Buse, K. Scattering-loss reduction of ridge waveguides by sidewall polishing. *Opt. Express* **2018**, *26*, 19815-19820, doi:10.1364/OE.26.019815.
30. Wu, R.; Zhang, J.; Yao, N.; Fang, W.; Qiao, L.; Chai, Z.; Lin, J.; Cheng, Y. Lithium niobate micro-disk resonators of quality factors above $10^7$. *Optics Lett* **2018**, *43*, 4116-4119, doi:10.1364/OL.43.004116.
31. Poberaj, G.; Hu, H.; Sohler, W.; Guenter, P. Lithium niobate on insulator (LNOI) for micro-photonic devices. *Laser Photon. Rev.* **2012**, *6*, 488-503, doi: 10.1002/lpor.201100035.
32. Sugioka, K.; Cheng, Y. Ultrafast lasers—reliable tools for advanced materials processing. *Light Sci. Appl.* **2014**, 3, e149, doi: 10.1038/lsa.2014.30.
33. Boes, A.; Corcoran, B.; Chang, L.; Bowers, J.; Mitchell, A. Status and potential of lithium niobate on insulator (LNOI) for photonic integrated circuits. *Laser Photon. Rev.* **2018**, 12, 1700256, doi: 10.1002/lpor.201700256.
34. Wolf, R.; Breunig, I.; Zappe, H.; Buse, K. Cascaded second-order optical nonlinearities in on-chip micro rings. *Opt. Express* **2017**, *25*, 29927-29933, doi:10.1364/OE.25.029927.

35. Wolf, R.; Breunig, I.; Zappe, H.; Buse, K. Q-factor enhancement of integrated lithium-niobate-on-insulator ridge waveguide whispering-gallery-mode resonators by surface polishing. In Proceedings of Laser Resonators, Microresonators, and Beam Control XIX; **2017**, p. 1009002, doi: 10.1117/12.2251516.
36. Guarino1, A.; Poberaj, G.; Rezzonico, D.; Degl'Innocenti, R.; Günter, P. Electro–optically tunable microring resonators in lithium niobate. *Nature Photon.* **2007**, *1*, 407-410, doi:10.1038/nphoton.2007.93.